\documentclass[traditabstract,preprint]{aa}
\usepackage{epsfig,graphicx,natbib}
\usepackage{longtable}
\usepackage{amssymb}

\def\PKS1830{\hbox{PKS\,1830$-$211}}
\def\kms{\hbox{km\,s$^{-1}$}}
\def\Tcmb{\hbox{$T_\mathrm{CMB}$}}

\begin{document}

\title{Probing the jet base of the blazar \PKS1830\ from the chromatic variability of its lensed images}
\subtitle{Serendipitous ALMA observations of a strong gamma-ray flare}
\author{I. Mart\'i-Vidal\inst{1} 
\and S. Muller\inst{1}
\and F.~Combes \inst{2}
\and S.~Aalto \inst{1}
\and A.~Beelen \inst{3}
\and J.~Darling \inst{4}
\and M.~Gu\'elin \inst{5,6}
\and C.~Henkel \inst{7,8}
\and C.~Horellou \inst{1}
\and J.~M. Marcaide \inst{9}
\and S.~Mart\'in \inst{10}
\and K.~M. Menten \inst{7}
\and Dinh-V-Trung \inst{11}
\and M.~Zwaan \inst{12}
}

\offprints{I. Mart\'i-Vidal, \email{mivan@chalmers.se}}
\institute{Department of Earth and Space Sciences, Chalmers University of Technology, Onsala Space Observatory, SE-43992 Onsala, Sweden
\and Observatoire de Paris, LERMA, CNRS, 61 Av. de l'Observatoire, 75014 Paris, France
\and Institut d'Astrophysique Spatiale, B\^at. 121, Universit\'e Paris-Sud, 91405 Orsay Cedex, France
\and Center for Astrophysics and Space Astronomy, Department of Astrophysical and Planetary Sciences, Box 389, University of Colorado, Boulder, CO 80309-0389, USA
\and Institut de Radioastronomie Millim\'etrique, 300, rue de la piscine, 38406 St Martin d'H\`eres, France 
\and Ecole Normale Sup\'erieure/LERMA, 24 rue Lhomond, 75005 Paris, France
\and Max-Planck-Institut f\"ur Radioastronomie, Auf dem H\"ugel 69, D-53121 Bonn, Germany
\and Astron. Dept., King Abdulaziz University, P.O. Box 80203, Jeddah, Saudi Arabia
\and Departamento de Astronom\'ia y Astrof\'isica, C/ Doctor Moliner 50, E-46100 Burjassot, Valencia, Spain
\and European Southern Observatory, Alonso de C\'ordova 3107, Vitacura, Casilla 19001, Santiago 19, Chile
\and Institute of Physics, Vietnam Academy of Science and Technology, 10 DaoTan, ThuLe, BaDinh, Hanoi, Vietnam
\and European Southern Observatory, Karl-Schwarzschild-Str. 2, 85748 Garching b. M\"unchen, Germany
}

\date{Accepted for publication in A\&A}
\titlerunning{Probing the jet base of \PKS1830}
\authorrunning{Mart\'i-Vidal et al. 2013}

\abstract{The launching mechanism of the jets of active galactic nuclei is observationally poorly constrained, due to the large distances to these objects and the very small scales (sub-parsec) involved. In order to better constrain theoretical models, it is especially important to get information from the region close to the physical base of the jet, where the plasma acceleration takes place. In this paper, we report multi-epoch and multi-frequency continuum observations of the $z$=2.5 blazar \PKS1830\ with ALMA, serendipitously coincident with a strong $\gamma$-ray flare reported by Fermi-LAT. The blazar is lensed by a foreground $z$=0.89 galaxy, with two bright images of the compact core separated by $1''$. Our ALMA observations individually resolve these two images (although not any of their substructures), and we study the change of their relative flux ratio with time (four epochs spread over nearly three times the time delay between the two lensed images) and frequency (between 350 and 1050~GHz, rest-frame of the blazar), during the $\gamma$-ray flare. In particular, we detect a remarkable frequency-dependent behaviour of the flux ratio, which implies the presence of a chromatic structure in the blazar (i.e., a core-shift effect). We rule out the possiblity of micro- and milli-lensing effects and propose instead a simple model of plasmon ejection in the blazar's jet to explain the time and frequency variability of the flux ratio. We suggest that \PKS1830\ is likely one of the best sources to probe the activity at the base of a blazar's jet at submillimeter wavelengths, thanks to the peculiar geometry of the system. The implications of the core-shift in absorption studies of the foreground $z$=0.89 galaxy (e.g., constraints on the cosmological variations of fundamental constants) are discussed.
} 

\maketitle

\section{Introduction}

Radio emission from the jets of Active Galactic Nuclei (AGN) has been extensively studied for more than 30 years. The early model of \cite{bla79} has been successfully used to explain, to-date, most of the AGN observations at several bands and spatial resolutions, from the radio to $\gamma$ rays (e.g., \citealt{beg84,mar92}). One of the main successes of this model in the radio band was the prediction of the so-called {\em core-shift} effect, i.e., the apparent shift of the core's position with frequency, due to optical depth effects. The effect was later discovered by \cite{mar84} and then studied in many AGN, from quasars and BL-Lacs (e.g., \citealt{kov08}) to low-luminosity AGN (e.g., \citealt{mar11}).

Although jets with constant opening angle (i.e., conical jets) can be used to model the spectra and the structures seen at VLBI scales (e.g., \citealt{lob98}), deviations from simple conical structures have been found (e.g., \citealt{asa12}). From the theoretical point of view, departures from a jet conical shape are expected, due to magneto-hydrodynamic collimation effects close to the jet base. \cite{mar80} built a parametric model of the continuum emission from AGN jets, from radio to X-rays, based on the earlier theoretical studies by \cite{bla74}, in which these collimation effects were taken into account. According to this model, the jet structure can be divided into three parts: 1) the jet ``nozzle'', that connects the central AGN engine (i.e., the super-massive black hole, SMBH) to the jet base; 2) the collimation region (with a concave shape), where the trajectories of the electrons are focused towards the jet direction, so part of the ``internal'' contribution to their Lorentz factors becomes a ``bulk'', or common, Lorentz factor; and 3) the free region (with a conical shape), where the trajectory of the plasma, once a maximum bulk Lorentz factor has been achieved, is believed to be ballistic. 

Despite the success of the standard jet model to explain the multi-band spectra and the VLBI structures seen in many radio-loud AGN, the injection and launching mechanisms of the jets are poorly understood. It is believed that the accretion of material into the SMBH triggers the injection of plasma into the jet. This leads to the well-known disc-jet connection, or the {\em fundamental plane} model of black-hole accretion (\citealt{mer03}). However, the exact mechanism from which the material is brought from the infalling region of the accretion disk to the base of the jet is unknown. Observational constraints on the emission from the regions involved in this process (e.g., \citealt{mar08b}) are essential for the progress of the theoretical models, although limited, due to the large distances to these objects and the small spatial scales (sub-parsec) involved.

In the present paper, we use ALMA continuum observations of the blazar \PKS1830\ to follow its variability at observing frequencies from 100 to 300\,GHz. The blazar is located at a redshift of $z$=2.507$\pm$0.002 (\citealt{lid99}) and is lensed by a foreground $z$=0.89 galaxy (\citealt{wik96}), which generates two bright and compact images of the core, separated by $1''$ and embedded in a weaker pseudo-Einstein ring seen at radio cm wavelengths (\citealt{jau91}). The compact images are located to the North-East and to the South-West of the pseudo-ring, and we hereafter refer to them as NE and SW images. Due to the steep spectral index of the pseudo-ring, only the NE and SW images remain visible at mm/submm wavelengths (with a lens magnification factor of about 5--6 for the NE image and 3--4 for the SW image, e.g., \citealt{nai93,win02}). The most precise measurement of the time delay between the two images is 27.1$\pm$0.6\,days (\citealt{bar11}, see also \citealt{lov98,wik01}), with the NE image leading. Thanks to this rare geometry, it is possible to measure the temporal and spectral variations of the flux ratio $\Re$ between the two images with high accuracy at submm wavelengths. We show that such observations can help constrain the physics of plasmon ejection in regions very close to (if not at) the base of the collimation region in the jet.

We adopt the cosmological parameters $H_0$=67.3\,\kms\,Mpc$^{-1}$, $\Omega_M$=0.315, $\Omega_\Lambda$=0.685 and a flat Universe (Planck collaboration; \citealt{ade13}). Accordingly, 1\,mas corresponds to 8.28\,pc at $z$=2.5 and to 8.00\,pc at $z$=0.89.

\section{Observations and data reduction}

The observations are part of an ALMA Early Science Cycle~0 project for the spectral study of absorption lines in the $z$=0.89 lensing galaxy toward \PKS1830 (Muller et al., in preparation). Here, we shall summarize briefly the points relevant for this paper. 

The observations were taken in spectral mode at frequencies around 100\,GHz (B3), 250\,GHz (B6), 290\,GHz (B7), and 300\,GHz (B7), targeting strong absorption lines of common interstellar species. The corresponding frequencies in the $z$=2.5 blazar rest-frame are $\sim$350, 880, 1020, and 1050\,GHz, respectively. For each tuning, four different 1.875\,GHz-wide spectral windows were set, each counting 3840 channels separated by 0.488\,kHz. Data were taken on 9--11 April (B6 and B7), 22--23 May (B3, B6, and B7), 4 June (B3 and B7), and 15 June 2012 (B3 and B6), see Table\,\ref{tab:cont-data}. The project was not designed as a monitoring of \PKS1830\ and Early Science observations were done under best effort basis from the ALMA observatory, hence the loose and irregular time sampling. The array configuration resulted in a synthesized beam of $\sim$\,$2''$ in B3, and $\sim$\,$0.5''$ in B6 and B7. The two compact images of the blazar (separated by $1''$) are easily resolved in the Fourier plane (see below), while their individual substructure (of mas scale) remains unresolved. 

The flux calibration was performed by short observations of Titan or Neptune. The absolute flux scale was set from a subset of short baselines, for which the planets were not resolved. This scaling was then bootstrapped to other sources for all baselines. We estimate an absolute flux accuracy of the order of $\sim$5\% in B3 and $\sim$10\% in B6 and B7.

VLA observations at 15 and 22.5\,GHz by \cite{sub90} reveal that the pseudo-Einstein ring has a steep spectral index ($\alpha$\,=\,1.5--2.0, with the flux $S\propto\nu^{-\alpha}$) compared to the two compact images ($\alpha$\,$\sim$0.7). Extrapolating the flux density of the components labelled C and D by \cite{sub90} to 100\,GHz and higher frequencies, we checked that their contribution becomes negligible for the ALMA observations. The continuum emission of the blazar images was thus modelled as two point sources (NE and SW). We used an in-house developed software ({\sc uvmultifit}; Mart\'i-Vidal et al., in preparation), based on the common-astronomy-software-applications (CASA) package \footnote{http://casa.nrao.edu/}, to perform the visibility model-fitting. 
Absorption lines from the $z$=0.89 galaxy and atmospheric lines were removed from the data before the fit. These lines are narrow ($\sim$100\,\kms, at most), compared to the bandwidth of each spectral window, and sparse, so that the remaining number of line-free channels was always large (typically $>$2000). The parameter uncertainties were derived from their covariance matrix, computed at the $\chi^ 2$ minimum, and scaled so that the reduced $\chi^2$ equals unity.

The estimated positions and fluxes of the NE and SW images were used to build a visibility model to perform phase self-calibration, with one complex gain solution every 30\,seconds. The high dynamic range of our observations ($\sim$1000--3000) and the large amount of antennas involved in the observations ($>$16) ensure that no spurious signal appears in the data after this self-calibration process (e.g., \citealt{marti08}). After self-calibration, the visiblity model-fitting process was repeated.

Rather than fitting the flux density of each of the two images, we have fitted the values $f_{NE}$, the flux density of the NE image, and $\Re$=$f_{NE}$/$f_{SW}$, their flux-density ratio. The fitting allows us to derive the ratio of flux densities between the images with high precision and accuracy. In particular, the uncertainty in the flux-ratio estimate is of the order of the inverse of the achieved dynamic range (i.e., $\sim 10^{-3}$) and, since we are comparing two sources within the same field of view (the field of view is much larger than the separation of $1''$ between the lensed images), it is free from systematics related to instrumental (e.g., bandpass) or observational (e.g., flux calibration) effects. We emphasize that, to the best of our knowledge, no flux-monitoring of any source has so far been reported at rest-frequencies between 350 to 1050\,GHz, with an accuracy similar to that achieved in the flux ratios derived from our ALMA observations toward \PKS1830.

\section{Results}

\begin{figure}[!th]
\centering
\includegraphics[height=9.6cm,angle=270]{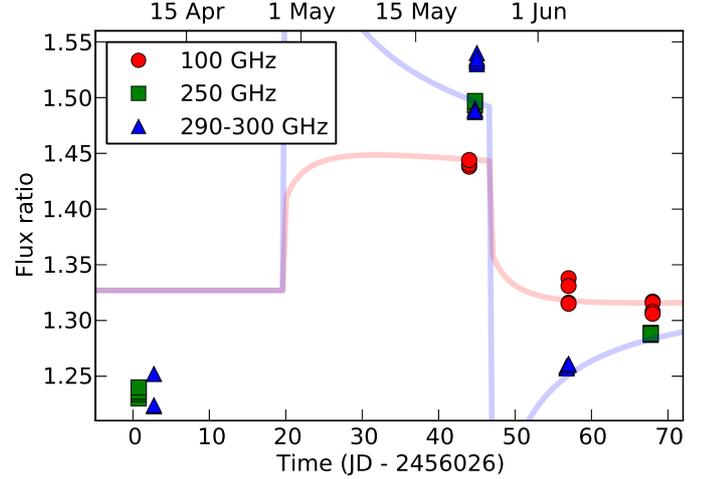}
\caption{Evolution of the flux-density ratio between the NE and the SW images, measured for each spectral window in our ALMA observations. The error bars are much smaller than the symbol sizes. The flux-ratio evolution based on our jet model (see text) is overlaid for frequencies of 100\,GHz (red) and 300\,GHz (blue). Note that at high frequencies (e.g., 300\,GHz), our model predicts fast and large variations of the flux ratio.}
\label{FigRatios}
\end{figure}

\begin{figure}[!th]
\centering
\includegraphics[width=9.6cm]{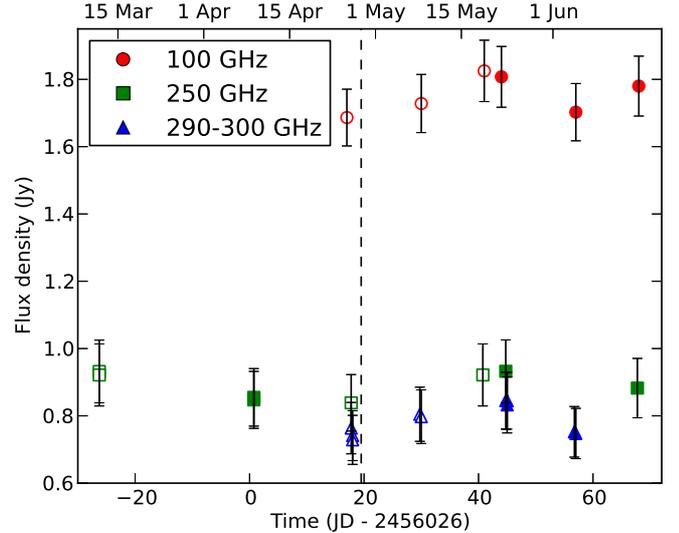}
\caption{Evolution of the submm flux density of the NE image of the blazar. Filled symbols are the actual flux density measurements of the NE image; empty symbols correspond to the flux densities of the SW image shifted backwards in time by 27 days (i.e., the time delay of the lens, \citealt{bar11}) and scaled up according to a quiescent flux ratio of $\Re_{quiet} = 1.34$ (see Table\,\ref{tab:parameters}). For each epoch and band, the flux densities of the four ALMA spectral windows (Table\,\ref{tab:cont-data}) were averaged together. The dashed line marks the time when the effect of the ejected plasmon begins to be seen in the NE image, according to our model.}
\label{FluxesFig}
\end{figure}

The time variations of the flux-density ratio $\Re$ between the two lensed images ($\Re$=$f_{NE}$/$f_{SW}$) during our ALMA observations are shown in Fig.\,\ref{FigRatios} for the different bands. Data cover a time span of roughly three times the time delay between the two lensed images. The flux ratio varies from a low value of $\Re \sim 1.2$ at the first data points on April 09--11 to a peak of $\Re \sim 1.5$ on May 22--23, and then decreases to $\Re \sim 1.3$ on June 04--15. The frequency dependence of the ratio is particularly interesting. While the data points do not show a large spread between bands in April, there is a strong frequency dependence during the flux-ratio increase in May, with higher ratios at higher frequencies, which becomes later inverted in the June data points (i.e., with higher ratios at {\em lower} frequencies). These strong changes in the flux ratio are not clearly reflected in the flux-density evolution of the blazar (Fig.\,\ref{FluxesFig}).

Besides the general time variations over a monthly timescale, we do observe rapid variations of $\Re$, of a few percent, over a much shorter timescale of hours (at 300\,GHz on May 23 and at 290\,GHz on April 11). This rapid behaviour seems to be seen at the high frequency end only, but the sparse time sampling of our observations does not allow us to investigate further the intra-day variability.

At cm wavelengths, \cite{rao88} and \cite{nai93} noted long ago that the flux ratio is varying with time and frequency.
A (sparse) monitoring of \PKS1830 at 3\,mm over a time period of 12\,years shows that the flux ratio can vary around a value of $\sim$1.6 (see \citealt{mul08}), with extreme excursions in the range 1--2. In these 3\,mm observations, however, the two images were not resolved and the ratio was determined from the saturation of the HCO$^+$ $J$=2-1 line at the velocity of the SW absorption, assuming a covering factor $f_c$ of unity. As $f_c$ is actually slightly lower than unity (Muller et al., in preparation), the flux ratio was likely slightly overestimated with this method. Using the BIMA interferometer at 3\,mm, \cite{swi01} resolved the two lensed images and could measure a flux ratio $\Re$ of 1.18$\pm$0.06 at the time of their observations (27-28 December 1999), within the range of ratios reported by \cite{mul08}. We should emphasize that there has been no previous measurement of the flux-ratio variability on timescales shorter than a day, that we are aware of.

\begin{table*}[ht] 
\caption{Flux densities of the NE image and flux ratios measured at the epochs of our ALMA observations.} \label{tab:cont-data} 
\begin{center} \begin{tabular}{ccccccc} \hline 
Band & Date & Julian day & Time  & Frequency $^\dagger$ & Flux of NE & Flux  \\ 
     &      & (-2456026.8) & (UTC) & (GHz)& image (Jy) $^\ddagger$   & Ratio \\ 
\hline 
        B6--250\,GHz &  09 Apr 2012 &  0.0 &         06:23--06:57 & 243.2 &  0.87(0.09) & 1.236(0.001)\\
 & & & & 245.1 &  0.86(0.09) & 1.238(0.001)\\
 & & & & 257.6 &  0.85(0.09) & 1.232(0.001)\\
 & & & & 260.5 &  0.84(0.08) & 1.230(0.001)\\
        B6--250\,GHz &  09 Apr 2012 &  0.1 &         07:42--08:16 & 243.2 &  0.87(0.09) & 1.234(0.001)\\
 & & & & 245.1 &  0.86(0.09) & 1.238(0.001)\\
 & & & & 257.6 &  0.83(0.08) & 1.236(0.001)\\
 & & & & 260.5 &  0.83(0.08) & 1.240(0.001)\\
        B7--290\,GHz &  11 Apr 2012 &  2.0 &         06:06--06:59 & 282.6 &  0.73(0.07) & 1.224(0.001)\\
 & & & & 284.4 &  0.73(0.07) & 1.224(0.001)\\
 & & & & 294.6 &  0.70(0.07) & 1.223(0.001)\\
 & & & & 296.4 &  0.71(0.07) & 1.224(0.001)\\
        B7--290\,GHz &  11 Apr 2012 &  2.1 &         07:47--08:40 & 282.6 &  0.79(0.08) & 1.251(0.002)\\
 & & & & 284.4 &  0.79(0.08) & 1.251(0.002)\\
 & & & & 294.6 &  0.77(0.08) & 1.252(0.001)\\
 & & & & 296.4 &  0.78(0.08) & 1.252(0.001)\\
        B3--100\,GHz &  22 May 2012 & 43.1 &         09:23--10:00 &  92.1 &  1.89(0.09) & 1.438(0.002)\\
 & & & &  94.0 &  1.86(0.09) & 1.440(0.002)\\
 & & & & 104.1 &  1.75(0.09) & 1.444(0.002)\\
 & & & & 106.0 &  1.73(0.09) & 1.444(0.002)\\
        B6--250\,GHz &  23 May 2012 & 43.9 &         04:38--05:14 & 243.3 &  0.96(0.10) & 1.493(0.001)\\
 & & & & 245.1 &  0.95(0.10) & 1.494(0.001)\\
 & & & & 257.6 &  0.91(0.09) & 1.497(0.001)\\
 & & & & 260.5 &  0.91(0.09) & 1.497(0.001)\\
        B7--300\,GHz &  23 May 2012 & 44.0 &         05:47--06:21 & 291.6 &  0.86(0.09) & 1.490(0.001)\\
 & & & & 293.5 &  0.85(0.08) & 1.488(0.001)\\
 & & & & 303.6 &  0.84(0.08) & 1.489(0.001)\\
 & & & & 305.5 &  0.83(0.08) & 1.487(0.001)\\
        B7--300\,GHz &  23 May 2012 & 44.1 &         09:14--09:51 & 291.6 &  0.86(0.09) & 1.530(0.002)\\
 & & & & 293.5 &  0.85(0.08) & 1.533(0.002)\\
 & & & & 303.6 &  0.84(0.08) & 1.534(0.002)\\
 & & & & 305.5 &  0.83(0.08) & 1.536(0.002)\\
        B7--300\,GHz &  23 May 2012 & 44.2 &         10:27--11:04 & 291.6 &  0.85(0.08) & 1.532(0.002)\\
 & & & & 293.5 &  0.83(0.08) & 1.535(0.002)\\
 & & & & 303.6 &  0.83(0.08) & 1.540(0.002)\\
 & & & & 305.5 &  0.82(0.08) & 1.540(0.002)\\
        B7--300\,GHz &  04 Jun 2012 & 56.0 &         07:18--07:52 & 291.6 &  0.77(0.08) & 1.257(0.001)\\
 & & & & 293.5 &  0.76(0.08) & 1.257(0.001)\\
 & & & & 303.6 &  0.74(0.07) & 1.258(0.001)\\
 & & & & 305.5 &  0.74(0.07) & 1.259(0.001)\\
        B7--300\,GHz &  04 Jun 2012 & 56.1 &         08:32--09:07 & 291.6 &  0.77(0.08) & 1.260(0.001)\\
 & & & & 293.5 &  0.77(0.08) & 1.261(0.001)\\
 & & & & 303.6 &  0.71(0.07) & 1.261(0.001)\\
 & & & & 305.5 &  0.74(0.07) & 1.261(0.001)\\
        B3--100\,GHz &  04 Jun 2012 & 56.1 &         09:42--10:18 &  92.1 &  1.79(0.09) & 1.338(0.002)\\
 & & & &  94.0 &  1.76(0.09) & 1.331(0.002)\\
 & & & & 104.1 &  1.64(0.08) & 1.316(0.002)\\
 & & & & 106.0 &  1.62(0.08) & 1.315(0.002)\\
        B6--250\,GHz &  15 Jun 2012 & 67.0 &         07:18--07:54 & 243.3 &  0.91(0.09) & 1.287(0.001)\\
 & & & & 245.1 &  0.91(0.09) & 1.289(0.001)\\
 & & & & 257.6 &  0.86(0.09) & 1.288(0.001)\\
 & & & & 260.5 &  0.85(0.09) & 1.288(0.001)\\
        B3--100\,GHz &  15 Jun 2012 & 67.1 &         08:53--09:29 &  92.1 &  1.86(0.09) & 1.317(0.002)\\
 & & & &  94.0 &  1.84(0.09) & 1.316(0.002)\\
 & & & & 104.1 &  1.72(0.09) & 1.308(0.001)\\
 & & & & 106.0 &  1.70(0.08) & 1.306(0.001)\\
\hline \end{tabular} 
\tablefoot{$^\dagger$  Frequency at the center of each ALMA spectral window. $^\ddagger$ We assume an absolute flux accuracy of 5\% at frequencies $\sim$100\,GHz (B3) and 10\% at frequencies between $\sim$200 and $\sim$300\,GHz (B6 and B7).} 
\end{center} \end{table*}

\begin{figure}[!th]
\centering
\includegraphics[width=9.0cm]{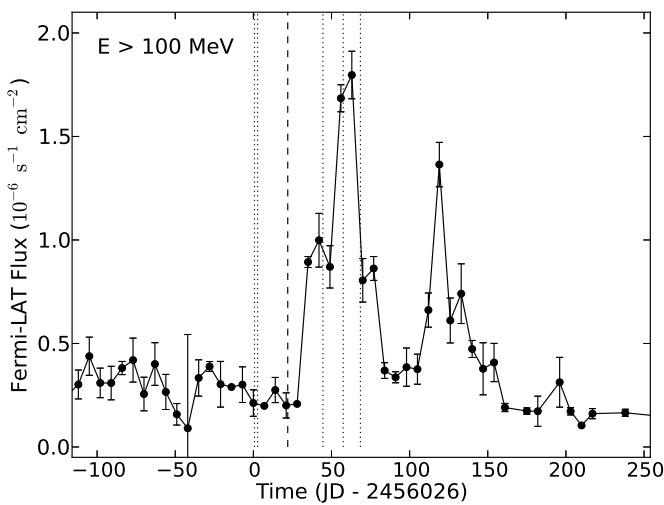}
\caption{Fermi-LAT light curve of \PKS1830. The dotted lines mark the epochs of our ALMA observations. The dashed line marks the time when the effect of the ejected plasmon begins to be seen at ALMA frequencies in the NE image, according to our model. Only Fermi-LAT points with a confidence level above 2\,$\sigma$ are shown. The time binning is of seven days.}
\label{FigFermi}
\end{figure}

PKS\,1830$-$211 is in the list of the Fermi Large Area Telescope (Fermi-LAT) monitored sources and its daily light-curve can be retrieved from the Fermi-LAT website \footnote{http://fermi.gsfc.nasa.gov/ssc/data/access/lat/msl\_lc/}. Several major $\gamma$-ray flares have been reported in the past (e.g., \citealt{cip10,cip12}), with amplitude variations up to a factor of tens on short timescales ($\sim$hours), revealing the strong intrinsic variation of the source (e.g., \citealt{don11}). The Fermi-LAT light curve (for energy above 100\,MeV) for year 2012, 
as retrieved from the Fermi-LAT public archive, is shown in Fig.\,\ref{FigFermi}. It can be seen from this figure, that our ALMA observations have been performed, although serendipitously, during the time of a major $\gamma$-ray flare, the strongest one in a period of two years, corresponding to an increase by a factor of up to seven within about one month. This coincidence provides us with a good opportunity to study the submillimeter counterpart of a $\gamma$-ray flare from a blazar.

Hereafter we discuss the interpretation of the temporal and chromatic evolution of the flux ratio in our ALMA data. We consider the potential effect of gravitational micro- and milli-lensing, showing that structural changes in the blazar's jet are needed anyway to explain the observations. Further, we consider a simple model of plasmon ejection, which can naturally and simply reproduce the flux-ratio evolution and its chromatic behaviour.

\section{Effects of micro- and milli-lensing?}

Micro- and milli-lensing events could introduce a variability in the flux ratio, but its chromatic changes directly imply a chromatic structure in the blazar (i.e., a core-shift effect). The variability in the amplification due to micro- and milli-lensing depends on the angular size of the source, $\theta_S$, relative to the typical angular size of the Einstein radius of the structure causing the light deflection, $\theta_E$. If the source is small compared to $\theta_E$, then the lensing variability can be large. Typically, an object in the lens plane with a mass $M$ can produce lensing variability if

\begin{equation}
\theta_S \lesssim \theta_E = \frac{1}{D_{OS}}\sqrt{\frac{4\,G\,M}{c^ 2}\frac{D_{LS}}{D_{OS}\,D_{OL}}},
\label{microEq}
\end{equation}

\noindent where $D_{IJ}$ is the angular distance from $J$ to $I$, and the subindices $O$, $L$, and $S$ stand for {\em observer}, {\em lens}, and {\em source}, respectively.

In the case of micro-lensing, e.g., by a stellar-mass object in the lens plane, we get a typical Einstein radius $\theta_E = 1.7$\,$\mu$as. This is a very small value compared to the expected angular size of the jet emission (e.g., \citealt{gea91}), falling by several orders of magnitude. Indeed, \cite{jin03} were able to slightly resolve the core size of \PKS1830 at 43\,GHz in their VLBA observations, getting diameters of $\sim0.5$\,mas (see their Fig.\,1). In the case of a conical jet, the size of the core emission should scale roughly as $\propto \nu^{-1}$ (for a concave jet, the dependence of size with frequency is weaker). Hence, a (conservative) estimate of the core size at our ALMA frequencies falls between 70\,$\mu$as (at 300\,GHz) and 215\,$\mu$as (at 100\,GHz), far too large to allow for an effective variability caused by micro-lensing.
We note that X-ray micro-lensing was suggested by \cite{osh01} to explain the large discrepancy between the intensity ratio at X-ray and the magnification ratio of the two lensed images in \PKS1830. The size of the blazar's X-ray emission region, i.e., of the order of a few Schwarzschild radii of a $\sim$\,$10^8$\,$M_\odot$ supermassive black hole (that is a few 10\,$\mu$pc), is indeed much smaller than that of the continuum-emitting region seen at ALMA frequencies.

Increasing the mass of the perturbing object in the lens plane by several orders of magnitude brings us to the milli-lensing regime. Here, the timescale for an apparent motion of an object in the $z$=0.89 galaxy is large: with a transverse velocity of 1000\,\kms, such an object would only cover an apparent drift of $\sim$0.5\,mpc (60 $\mu$as) within one year (observer frame). Therefore, the timescale of milli-lensing would be too long to explain the short timescales observed in the flux-ratio evolution (i.e., days or a few weeks).

On the other hand, a plasmon travelling at near speed of light in the blazar's jet would cover an apparent projected distance of $\sim$0.2\,mas in the lens plane within one month, so that milli-lensing could not be formally ruled out to explain variabilities of the order of weeks or a few months. 
However, the intra-day variability detected in our ALMA observations cannot be explained with milli-lensing. In any case, we emphasize that an intrinsic variability in the blazar's jet is required for milli-lensing to work on timescales of the order of less than one year.

\section{Intrinsic variability in the blazar}

The most simple way to explain the temporal and chromatic changes in the flux ratio is to consider intrinsic variability in the jet of the blazar, which must indeed be variable by nature. The odd behaviour in the evolution of the flux ratio, $\Re$, 
can be explained using a simple model of an over-density region (plasmon), travelling downstream the jet. From the evolution of the flux-density of the NE image, we can set an upper bound to the flux-density increase of only 5\% at 100\,GHz (observer-frame) during the flare (see Fig.\,\ref{FluxesFig}). We note that this is probably the weakest flaring event ever detected from a blazar at submm wavelengths. In contrast, the $\gamma$-ray emission shows a much larger variability, of a factor of up to seven during the same period.

\subsection{Model of the blazar's jet}

From 43\,GHz VLBA observations, \cite{gar97} and \cite{jin03} reported structural and temporal variations in the radio core/knot images of \PKS1830. \cite{jin03} could measure changes in the relative distance between the centroids of emission of the NE and SW images to up to 200\,$\mu$as in a few months. These changes were interpreted by \cite{nai05} as due to a helical jet, with a jet precession period of about one year (corresponding to an intrinsic period of $\sim$30\,yr for the source at $z$=2.5), possibly due to the presence of a binary black-hole system at the center of the AGN. 
Morphological changes in the continuum emission are also believed to be responsible for the time variations observed in the $z$=0.89 molecular absorption (\citealt{mul08}).

As far as we know, there has been no further attempt to observe the evolution of \PKS1830's core/jet structure at high angular (VLBI) resolution since the observations reported by \cite{jin03}. We show below that our ALMA data can help shed new lights on this interesting system.

\begin{figure}[!th]
\centering
\includegraphics[width=8cm]{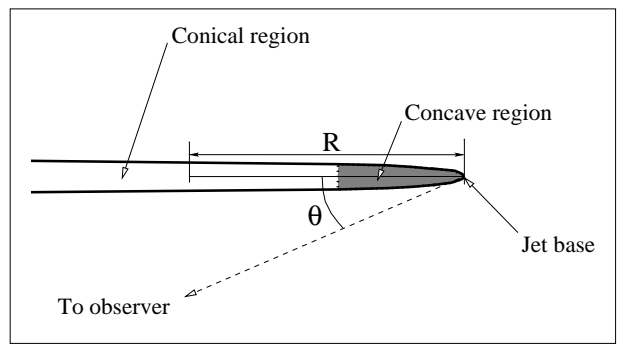}
\caption{Sketch of the path followed by the jet plasma during our observations. Not to scale. The concave region (gray) should be much smaller than the conical region (white). Moreover, the line of sight coincides with the precession axis of the jet tube. The precession angle is $\theta$.}
\label{SketchJet}
\end{figure}

The geometry of the blazar is illustrated as a sketch in Fig.\,\ref{SketchJet}. Based on the model of \cite{nai05}, we set a jet viewing angle of $\theta = 3^{\circ}$, and assume that the precession axis almost coincides with the viewing direction (so that the viewing angle does not change with time). Indeed, the time span of our observations is much shorter than the precession period reported by \cite{nai05}, so that we can consider the jet viewing angle as a constant in any case.

By fixing the viewing angle, we can perform numerical modelling of the mm-submm emission from the jet, using the parametric model by \cite{mar80}. The details of our implementation of the Marscher's model are described in Appendix \ref{ModelApp}. We have simulated a flare in the jet as due to a narrow over-density in the population of synchrotron-emitting particles (hereafter, a {\em plasmon}), due to either a sudden increase in the accretion rate by the SMBH (i.e., related to the disk-jet connexion) or triggered by an internal shock in the jet (e.g., \citealt{mim04}). Although elaborated hydrodynamical codes are used to model the propagation of internal shocks in jets (e.g., \citealt{mim04}, \citealt{bot10}), we have built a simplified model for the evolution of the jet flare (see Appendix \ref{ModelApp}), using a minimum number of model parameters in compromise with our limited amount of observations. Then, using a time delay of 27~days between the NE and SW images (the NE image leading), we have computed the flux-density ratio as a function of time and observing frequency. A direct comparison between the observed ratios and the model predictions allows us to constrain the defining parameters of the jet model by means of least-squares minimization. 

Our jet model depends on several parameters, which are listed in Table\,\ref{tab:parameters} and described in Appendix\,\ref{ModelApp}. We distinguish between two kinds of fitting parameters. The first kind describes the quiescent state of the jet: on the one hand, we have the power index of the electron energy distribution, $\gamma$, and the bulk Lorentz factor of the electrons, $\Gamma$; on the other hand, the opacity, $\tau_{\nu_0}$, for a given reference frequency, $\nu_0$, and distance to the jet base, $R_0$. Finally, the integrated flux density at the reference frequency over the jet, $F_{\nu_0}$, and the flux ratio of the NE image to the SW image in the quiescent state, $\Re_{quiet}$. For the case of a concave jet, we must add the curvature index of the jet surface, $\beta$.

The second kind of parameters describes the flare as due to an over-density of plasma, which travels through the jet with the same local bulk Lorentz factor as that of the quiescent plasma. The parameters used here are the width of the over-density region, $\Delta$, the density contrast factor, $K$, and the time of injection of the plasmon into the jet, $t_0$. An increase in the local magnetic field is also applied, to keep a constant ratio of the particle and field energy densities.

Regarding the fixed parameters in the model, we have the jet viewing angle, $\theta$ (fixed to $3^ {\circ}$) 
and the time delay of the lens, $\Delta \tau$ (fixed to 27\,days). We notice that variations in these fixed parameters within reasonable limits (0.5 degrees for $\theta$ and a few days for $\Delta \tau$) do not change the conclusions reported in the following sections (a small change in $\Delta \tau$ basically translates into a change in $t_0$).

\begin{table*}[ht]
\caption{Parameters of our jet model.}
\label{tab:parameters}
\begin{center}
\begin{tabular}{llll}
\hline
\multicolumn{1}{c}{Parameter} & \multicolumn{1}{c}{Value (concave)} & \multicolumn{1}{c}{Value (conical)} & \multicolumn{1}{c}{Notes}\\
\hline
\multicolumn{3}{c}{\em Fixed} & \\
Jet viewing angle, $\theta$                              & 3 deg.             & 3 deg. & \cite{nai05} \\
Time delay, $\Delta \tau$, between the NE and SW images     & 27\,days           & 27\,days & \cite{bar11}) \\
\hline
\multicolumn{3}{c}{\em Quiescent state} & \\
Power index of e$^-$ energy distribution, $\gamma$       & $2.8 - 3.0$        & $1.0 - 1.2$ &\\
Reference distance to the jet base, $R_0$                &  $1.05 - 1.27$\,pc &  $0.88 - 1.11$\,pc &\\
Bulk Lorentz factor, $\Gamma_0$, at $R_0$                & $9.1 - 12.2$        & $7.8 - 8.7$ &\\
Reference frequency, $\nu_0$                             & 100\,GHz           & 100\,GHz &\\
Opacity, $\tau_{\nu 0}$, at $\nu_0$ and $R_0$            & 1                  & 1 &\\
Total flux density, $F_{\nu_0}$, at $\nu_0$              & $1.73 - 1.78$\,Jy  & $0.9-1.1$\,Jy &\\
Flux-density ratio, $\Re_{quiet}$=$F_{NE}$/$F_{SW}$      & $1.325 - 1.350$    & $1.350 - 1.370$ &\\
Curvature index of the jet surface, $\beta$              & $0.039 - 0.087$    & $-$ & \\
\hline
\multicolumn{3}{c}{\em Flare} & \\
Width of the over-density region, $\Delta$               & $1.7 - 2.0$\,mpc   & $1.0 - 1.3$\,mpc & \\
Density contrast factor, $K$                             & $200 - 210$        & $110 - 140$   & \\
Injection time of plasmon, $t_0$ (days after 1st epoch)  & $19.2 - 19.5$      & $18.8 - 19.1$ & \\
\hline
\end{tabular} 
\end{center} 
\end{table*}

The set of parameters for the quiescent and flaring stages are listed in Table\,\ref{tab:parameters}.
We note that the large number of parameters to be fitted (8 for a conical jet, 9 for a concave jet), together with the limited amount of data and the non-linearity in the behaviour of most of the parameters, makes it difficult to constrain the parameter space in a statistically robust way. Thus, instead of one single estimate for each parameter, we explored the parameter space and report the range of values that match the data with a similar quality (maximum increase in the $\chi^ 2$ of 20\% with respect to the minimum value). 

\subsection{Fit to a conical jet}

\begin{figure*}[!ht]
\centering
\includegraphics[width=18cm]{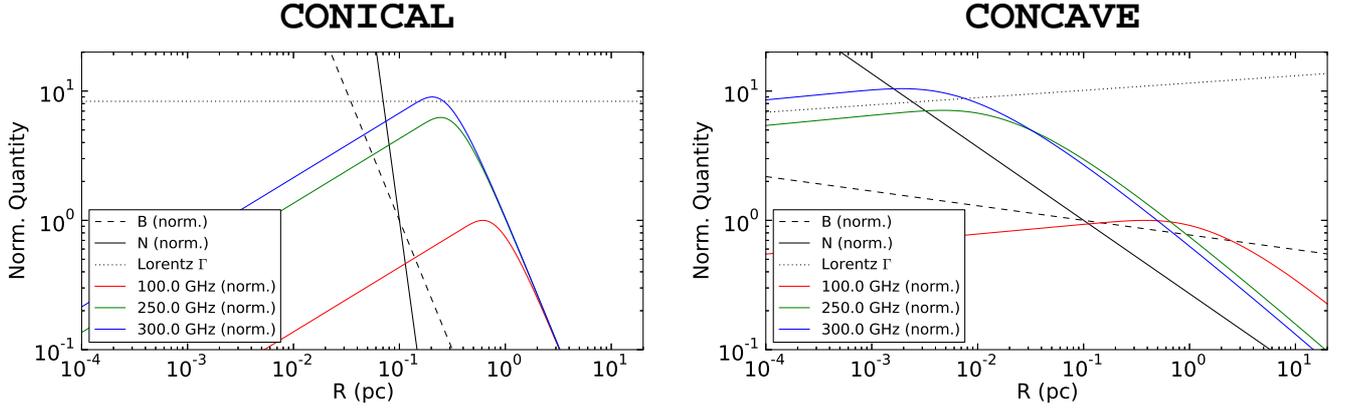}
\caption{Quantities derived from our simplified jet model. The magnetic field $B$ and electron density $N$ (dashed and solid black lines) are shown as a function of radial distance from the jet origin, normalized to their values at $10^{-1}$\,pc; the Lorentz factor, $\Gamma$, (dotted line) is shown unnormalized; the jet brightness at 100, 250, and 300\,GHz (blue, green, and red lines) is shown normalized to the brightness peak at 100\,GHz. Left, a fitting model obtained assuming that the emission comes from the conical (i.e., {\em free}) jet region. Right, a fitting model assuming that the emission comes from the concave jet region. See text for details.}
\label{SketchFig}
\end{figure*}

\begin{figure*}[!ht]
\centering
\includegraphics[width=18cm]{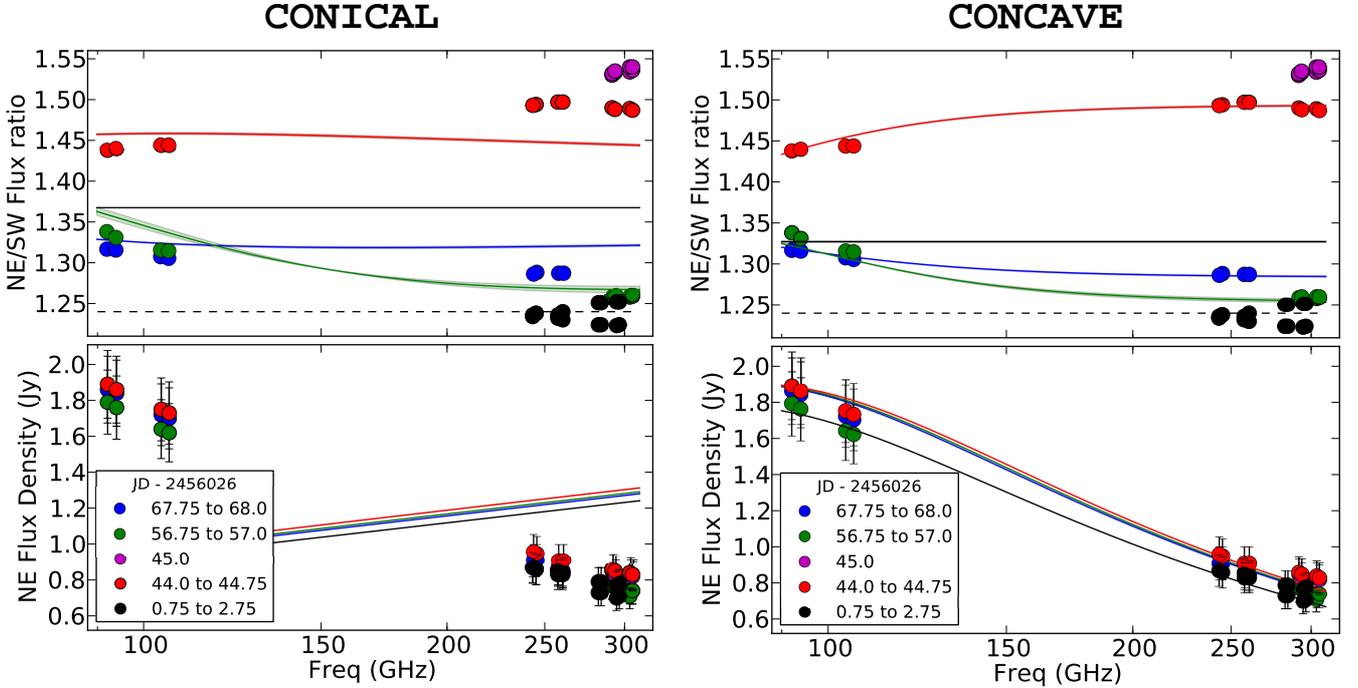}
\caption{Top, flux-density ratio of the NE image to the SW image. Bottom, flux density of the NE image. Circles are the ALMA data and lines are the predictions from our best-fit jet model. Different colors correspond to different observing epochs. Left, fit to the conical-jet model. Right, fit to the concave-jet model. The dashed line marks the lower flux-density ratio expected from an (unconstrained) previous flare (see text for details).}
\label{FitPlotFig}
\end{figure*}

We show in Fig.\,\ref{SketchFig} (left) the behaviour of the magnetic field, particle density, bulk Lorentz factor, and flux-density per unit length of a conical-jet model fitted to our ALMA data. The values of the parameters used are given in Table\,\ref{tab:parameters}. The plot of model versus data is shown in Fig.\,\ref{FitPlotFig} (top left). The general trend of the flux-density ratios is roughly followed by the model, with higher ratios in the May epochs and lower ratios in the June epochs. The frequency dependence of the ratios in the June epochs is also recovered. Regarding the spectrum (Fig.\,\ref{FitPlotFig}, bottom left), the conical-jet model is clearly unable to reproduce the optically-thin spectrum seen in the data ($\alpha$=0.7). Indeed, nearly-flat spectra are expected from conical jets in energy equipartition between the leptons and the magnetic field. This effect, known as the ``cosmic conspiracy'', is due to the particular dependence of all these quantities with distance to the jet base (\citealt{mar77,bla79}). However, this is only true as long as the jet base is opaque to the radio emission. If the frequencies are high enough, the whole jet becomes optically thin, and the spectrum steepens (e.g. \citealt{mar80}). The peak emission at these high frequencies is obviously located close to (it not at) the physical base of the jet, where the conical model does not apply. Hence, the frequency range with a steep spectrum, which implies a (nearly) optically-thin jet, shall be modelled using a concave jet, as we describe in the next section. 

According to our model, the epochs on April (black color; Fig.\,\ref{FitPlotFig}, left) were taken well before the flare. Then, the epochs in May (red color) were taken while the flare had already arrived to the NE image (hence the larger flux-density ratios). Finally, the epochs on June (green and blue colors) were taken when the flare already arrived to the SW image, hence explaining the lower flux-density ratios {\em and} the slightly higher ratios at the lower frequencies (i.e., those frequencies for which the flare was not strongly illuminating the SW image yet).

We must notice that the model prediction of flux-density ratios for the data taken on April falls above the data (Fig.\,\ref{FitPlotFig}, top). An earlier flare in its final stage (i.e., illuminating only the SW image) is needed to explain the lower ratios at these epochs. Unfortunately, the lack of earlier data prevents us of constraining any quantity for this eventual previous flare.

\subsection{Fit to a concave jet}

The quantities related to a concave-jet model are shown in Fig.\,\ref{SketchFig} (right). We notice the increase in the bulk Lorentz factor with distance, as well as the slower decrease in magnetic-field strength and particle density, which steepen the synthesized spectrum. The parameters used to generate this plot are shown in Table\,\ref{tab:parameters}. The peak intensity at high frequencies is located at a distance very close to the jet origin (a few mpc). If this distance would be of the order of the nozzle size (e.g., similar to the case of 3C\,345, \citealt{mar80}), Fig.\,\ref{SketchFig} (right) would suggest that the jet is almost (if not completely) optically-thin to the emission at our highest frequencies.
This would, indeed, steepen the observed spectrum (as it is discussed in the previous section).

The plot of observed ratios vs. model is shown in Fig.\,\ref{FitPlotFig} (top right). The fit to the May epochs is improved and the model can now reproduce the spectrum (Fig.\,\ref{FitPlotFig}; bottom right). Same as for the conical-jet model, an earlier flare is needed to explain the lower flux ratios observed in April.

\begin{figure}[!ht]
\centering
\includegraphics[width=5cm,angle=270]{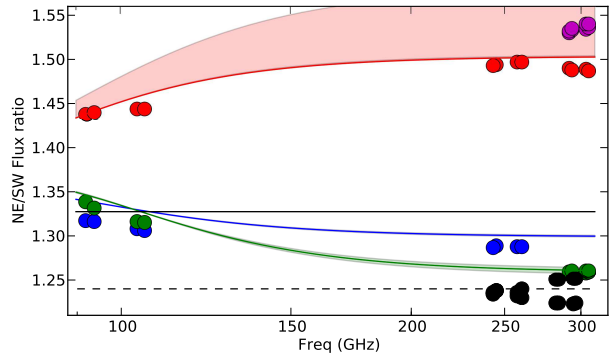}
\caption{Fit of a concave-jet model to the data, but adding a second (and weaker) flare, emitted after the first one. Same color code as in Fig.\,\ref{FitPlotFig}. The shaded area covers the variability of the model within $\pm$1\,day. Notice the large variability of the model on 22--23 May.}
\label{NewFitConcave}
\end{figure}

We notice that some flux ratios in May at 300\,GHz (Fig.\,\ref{FitPlotFig}, magenta) cannot be reproduced by the model. A rapid variability is needed in this case, since the observed ratios changed from 1.49 to 1.54 in only $\sim$5\,hours.
Nevertheless, it is remarkable that our simple model is able to reproduce the time scale in the evolution of the flux ratios in all the other cases. Indeed, the larger 300\,GHz ratios in May could be explained by substructure in the plasmon or successive flaring events. To illustrate this, we show in Fig.\,\ref{NewFitConcave} the results of a model with a concave jet, adding an extra flare, 10 times weaker than the first flare and emitted $\sim$22\,days later. This new model is able to predict a rapid variability for the epochs in May. 
Unfortunately, we do not have enough observations to perform a (statistically meaningful) fit of models more complicated than one simple plasmon.

\subsection{Constraints on the jet physics} \label{ConstrSec}

As described in Appendix \ref{ModelApp}, our model assumes a jet with a very small opening angle. Indeed, our fitting parameters determine all the proportionality constants between the distance to the jet base, $R$, the source function, $\epsilon_\nu/\kappa_\nu$, and the opacity, $\tau_\nu$, without any need of using an absolute width of the jet. Hence, the magnetic field and particle density cannot be directly determined from our fitted parameters, unless we assume a given absolute size (i.e., basically, an opening angle) of the jet tube.

\begin{figure*}[ht!]
\centering
\includegraphics[width=18cm]{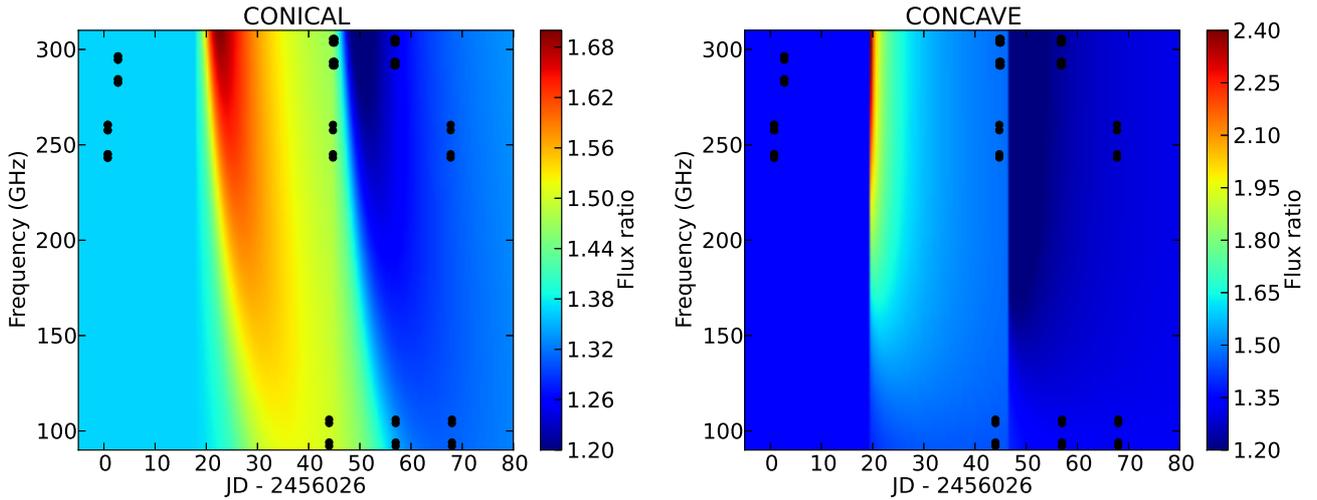}
\caption{Flux ratios derived from our jet model, as a function of time and frequency. Left, using a conical jet. Right, using a concave jet. Black points correspond to the epochs and frequencies of our ALMA observations.}
\label{ModelFig}
\end{figure*}

Nevertheless, a quantity that can be well constrained in our model is the {\em core-shift} among the observing frequencies (i.e., roughly speaking, the separation between the $\tau=1$ surfaces at the different frequencies; \citealt{bla79}). Even though the resolution of our ALMA observations is not high enough to actually {\em measure} the core-shift, we can still estimate it indirectly based on our model. For any pair of frequencies, the core-shift is related to the time needed by the plasmon to travel from one $\tau=1$ surface to the other. Since the speed of the plasmon is likely close to the speed of light, the time scale in the variability of the flux-density ratios (Fig.\,\ref{FitPlotFig}, top) constrains the distance between the cores at our three observing frequencies. We notice, though, that the dependence of the opacity with distance to the jet base is very smooth in the concave-jet model, so the effect of the core-shift in this model is less pronounced than in the conical-jet model. This can be easily seen in Fig.\,\ref{ModelFig}, where the complete simulation of the flux-density ratios is shown for both, the conical-jet and the concave-jet models. 
However, both models still allow us to estimate the core-shift from the observed time evolution of the flux-density ratios. From our jet models, both conical and concave, we estimate a distance of 0.3--0.5\,pc between the cores at 100\,GHz and 300\,GHz. Assuming a viewing angle of $3^{\circ}$ for the jet, the distance between the cores translates into a projected angular shift of 2--3\,$\mu$as. However, the blazar is lensed, and a magnification of 3--6 (depending on the image) eventually results in an apparent core-shift of 5--8\,$\mu$as between 100 and 300\,GHz (for a shift of 0.5\,pc)\footnote{The size amplification goes as the square root of the magnification factor.}.

Concerning the over-density region of emitting particles in the jet (i.e., the plasmon), satisfactory fits are only obtained when the density contrast is high ($>$100) and the size is narrow ($\sim$1\,mpc). Since the length of the radio jet (pc scale) is much larger than the size of the plasmon, the contribution of the latter to the total submm emission is small. This would explain the weak flux-density enhancement at submm wavelengths, due to flare {\em dilution}. In contrast, if the $\gamma$-ray emitting region is small compared to the radio jet (e.g., \citealt{val96}), the $\gamma$-ray flare would not be significantly diluted, and the $\gamma$-ray variability would be large compared to that at submm wavelengths, as observed (Figs.\,\ref{FluxesFig} and \ref{FigFermi}). 

Regarding the injection time of the plasmon, we estimate it to be about 20 days after the first ALMA observation. However, this is not a direct estimate (based on the evolution of the flux density), but it is based on our model and on the time delay of the lens. In any case, we do not expect the real injection time to differ from our estimate by more than a few days, as we discuss in the following lines. The chromatic behaviour seen in the flux ratios observed on May 23 and June 6 (Fig.\,\ref{FigRatios}) {\em must} be due to the flare reaching the SW image between these two epochs \footnote{The $\gamma$-ray light curve shows an enhancement of the flare emission in the form of a second peak (i.e., JD around 2456090; Fig. \ref{FigFermi}). The ratio of the two $\gamma$-ray peaks (i.e., the first one around JD 2456066), is consistent with the magnification ratio between the two lensed images. This is further evidence that the flare is due to intrinsic variability in the blazar.}.
Since the time delay is well known, our model allows us to constrain the start of the flare in the NE image with a precision of just a few days. Thus, the time-lag between the $\gamma$-ray and submm flares should not be larger than a few days (Fig.\,\ref{FigFermi}; dashed line), suggesting that both flares originate at the same region in the jet.
 The co-spatiality of the $\gamma$-ray and submm flares is a direct prediction of the shock-in-jet model (\citealt{val96}), in which the $\gamma$-rays would be created by Compton up-scattering in the region of synchrotron-emitting electrons. Similar short time-lags between $\gamma$-ray and mm/submm flares have been seen in other blazars, although the flaring activity can sometimes be located at distances as large as several parsecs from the central engine (e.g., \citealt{agudo}). In our case, the flaring event should be generated at, or close to, the base of the jet (i.e., at the region where the submm emission changes from optically-thick to optically-thin), to account for the observed frequency-dependent evolution of the image flux-ratios.

A few months after our ALMA observations, another $\gamma$-ray flare is seen in the Fermi-LAT light curve (i.e., between JD 2456140 and 2456180). Since the $\gamma$-ray flaring events in \PKS1830\ are rare, the overall $\gamma$ activity in 2012 (concentrated within a few months) may be related. Indeed, similar multiple $\gamma$-ray flares have been observed in other blazars. For example, \cite{ori13} observed a double $\gamma$-ray flaring event in PKS\,1510$-$089, where the first episode was not seen in radio (suggesting an origin close to the radio-opaque base of the jet), but the second episode had a strong radio counterpart, indicating an origin several parsecs downstream the jet. Similarly, the consecutive $\gamma$-ray flaring events seen in \PKS1830\ in 2012 might be the signature of the same plasmon, propagating downstream the jet.

\subsection{Comparison with mm/submm flaring activity in other AGN}

There is an intense observational work in flux-density monitoring of blazars, covering different bands of the electromagnetic spectrum (e.g., \citealt{gio12}; \citealt{kur13}, and references therein). Nevertheless, the study of blazar variability at mm and submm wavelengths is technically limited, so only strong flares observed in bright and/or nearby sources can be detected (e.g., \citealt{gio12}). Even with this limitation, the detection of flares at mm wavelengths, lasting several weeks, is not rare in sources as instrinsically weak as Sgr A* (\citealt{miy06}).

The intensity of the flare reported in the present paper is much weaker than the quiescent flux density of the blazar's jet ($\sim$5\% at 100\,GHz). This flaring event  could only be spotted from the flux-ratio evolution, thanks to the time delay between the two lensed images. On the one hand, relative fluxes among images are free of absolute calibration effects, so very weak flares can be clearly detected (variabilities as weak as a few times the inverse of the dynamic range in the image can be identified). On the other hand, the large frequency coverage of our observations (a factor 3 in frequency space) allows us to see the frequency dependence in the jet emission as the plasmon travels through it. Hence, the possibility of monitoring the evolution of very weak flares (and at very different ALMA frequencies), using the flux-density ratios in the \PKS1830 images, opens a new window for the study of blazar variability at frequencies and energy regimes that are not explored yet.

\section{Implications for the absorption studies in the foreground $z$=0.89 galaxy}

In general, the effects of the core-shift and the frequency-dependent size of the continuum emission should be taken into account as possible sources of systematics in absorption studies. Different continuum emission (at different frequencies) would illuminate different regions of the absorbing molecular gas in the foreground galaxy. As we discuss in Sect.\,\ref{ConstrSec}, a core-shift of 5\,$\mu$as (for the SW image, with a magnification factor of 3) can be expected between 100 and 300\,GHz. Projected in the plane of the foreground $z$=0.89 galaxy, this value translates into a distance of $\sim$0.04\,pc. At frequencies lower than those of our ALMA observations, the effect can be larger. Using our model, we can estimate the core-shift in the blazar's jet at any pair of frequencies (e.g., \citealt{lob98}) using:

\begin{equation}
\Delta\alpha = \Omega \left( \frac{1}{\nu_2} - \frac{1}{\nu_1} \right),
\label{CoreShiftEq}
\end{equation}

\noindent where $\nu_1$ and $\nu_2$ are the observing frequencies and $\Omega$ is the normalized core-shift. Energy equipartition (or a constant ratio) between particles and fields is assumed. This equation also assumes a conical jet for the emission at all frequencies, but it can still be used as a rough approximation using the core-shifts determined at the ALMA frequencies. As can be seen from Eq.\,\ref{CoreShiftEq}, the core-shift increases rapidly with decreasing frequency. Our model suggests a value of $\Omega$\,$\sim$0.8\,mas\,GHz (this value already includes a fiducial lens magnification of three, i.e. for the SW image). 

\cite{bag13} used several methanol lines redshifted between 6 and 32\,GHz to constrain a cosmological variation of the proton-to-electron mass ratio, $\mu$, at $z$=0.89 toward \PKS1830. Their method lies on the fact that a cosmological variation of $\mu$ would introduce velocity shifts between different lines of methanol (\citealt{jan11}). For their observing frequencies, our estimate of the core-shift is of the order of 0.1\,mas, which corresponds to a projected linear scale in the foreground $z$=0.89 galaxy of $\sim$1\,pc, comparable to the typical size of molecular clumps.
Hence, the different methanol lines might trace gas with a slightly different kinematics, introducing a systematic source of uncertainty on velocity offsets and on a constraint on $\mu$ variation. According to the Larson law (\citealt{lar81}), clumps of 1\,pc would have a velocity dispersion of a few \kms. In turn, an offset of 1\,\kms would translate into an uncertainty of $\sim$10$^{-7}$ in the estimate of $\Delta\mu/\mu$.

The molecular absorption toward \PKS1830\ can also be used to measure the temperature of the cosmic microwave background, \Tcmb, at $z$=0.89. For this purpose, several molecular transitions, in general at different frequencies, need to be observed to derive the excitation conditions of the gas. \cite{sat13} made milli-arcsecond resolution Very Long Baseline Array observations of the HC$_3$N $J$=3-2 and 5-4 transitions redshifted to 14.5 and 24.1\,GHz, respectively. An excitation analysis based on their lower (26\,mas) resolution images yielded a value \Tcmb=$5.6^{+2.5}_{-0.9}$\,K, consistent with value predicted by the standard cosmology (\Tcmb=5.14\,K at $z$=0.89). However, their full resolution data yielded significantly lower values of 1--2.5\,K. As possible explanations of this finding, \cite{sat13} discussed both of the latter scenarios, illumination of different absorbing gas volumes and the core-shift effect, which would amount to a displacement of $\sim$0.022\,mas (or $\sim$0.2\,pc) at their observing frequencies. In contrast, \cite{mul13} found a value \Tcmb=(5.08$\pm$0.10)\,K from a variety of molecules observed between 7\,mm and 3\,mm with the Australia Telescope Compact Array, (at epochs with no apparent $\gamma$-ray flaring activity). In that study, the effect of the core-shift on the determination of \Tcmb\ is minimized by the use of higher frequencies and smeared out by the use of multiple frequency combinations in the excitation analysis.

Future multi-frequency VLBI observations will be needed to address the issue of the frequency-dependent continuum illumination for absorption studies.

\section{Summary and conclusions}

We present multi-epoch and multi-frequency ALMA Early Science Cycle~0 submm continuum data of the lensed blazar \PKS1830, serendipitously coincident with a strong $\gamma$-ray flare observed by Fermi-LAT. The ALMA observations, spanning a frequency range between 350 and 1050\,GHz in the $z$=2.5 blazar rest-frame, resolve the two compact lensed images of the core of \PKS1830. This allows us to monitor the variation of their flux-density ratio as a function of time and frequency during the $\gamma$-ray flare. 

The time variations are large ($\sim$30\%) and, even more interestingly, show a remarkable chromatic behaviour. We rule out the possibility of micro- and milli-lensing, based on the timescale of the variability. Instead, we propose a simple model of jet and plasmon which can explain naturally the time evolution and frequency dependence of the flux ratio. This picture is consistent with the $\gamma$-ray flaring activity. According to the model, the frequency-dependence of the flux ratio is related to opacity effects close to the base of the jet. Since the time-lag between the $\gamma$-ray and the submm flares is short (a few days at most), we suggest that both flares are co-spatial, in agreement with the shock-in-jet model of $\gamma$-ray emission.

The frequency-dependence of the flux ratio is a direct probe of the chromatic structure of the jet, implying the existence of a core-shift effect in \PKS1830\ blazar's jet (just as seen in many other AGN jets). This core-shift should be considered as a possible source of systematics for absorption studies in the foreground $z$=0.89 galaxy, as the line of sight through the absorbing gas varies with the observing frequency.

Given the peculiar properties of \PKS1830\ at submm wavelengths (resolvability of the lensed images, high radio brightness, and achievable accuracy of the flux-ratio measurements), we suggest that \PKS1830\ is probably one of the best sources (if not the best) for future monitoring of (even weak) submm variability, related to activity at the jet base of a blazar, and study of the radio/$\gamma$-ray connection.

\begin{acknowledgements}
This paper makes use of the following ALMA data: ADS/JAO.ALMA\#2011.0.00405.S . ALMA is a partnership of ESO (representing its member states), NSF (USA) and NINS (Japan), together with NRC (Canada) and NSC and ASIAA (Taiwan), in cooperation with the Republic of Chile. The Joint ALMA Observatory is operated by ESO, AUI/NRAO and NAOJ. The financial support to Dinh-V-Trung from Vietnam's National Foundation for Science and Technology (NAFOSTED) under contract 103.08-2010.26 is greatly acknowledged. Data from the Fermi-LAT public archive were used. We thank the ALMA scheduler for having executed the observations, by chance, right at the time of the $\gamma$-ray flare. 
\end{acknowledgements}

\appendix

\section{Simplified jet model} \label{ModelApp}

Following \cite{mar80}, a simple model of a radio-loud AGN jet can be divided into three parts. The first one, the ``nozzle'', connects the central engine (SMBH) to the physical base of the jet; the second part consists of a small region in the jet base, the launching region, where the electrons of the plasma are accelerated to relativistic speeds; the third part, the conical jet, corresponds to the region usually observed, and resolved, in VLBI at mm and cm wavelengths. In this conical region, the bulk Lorentz factor of the plasma has reached a maximum value, and the only energy gain of the leptons is due to synchrotron self-absorption (SSA), which is an effect much smaller than the energy loss due to both, expansion and synchrotron radiation.

The jet diameter, $r$, is parameterized in the conical region as a function of the distance to the central engine, $R$, in the form $r \propto R$. Regarding the particle density, it follows a power-law of energy, $N = N_0 E^{-\gamma}$, up to a cutoff energy $E_{max}$. The factor $N_0$ decreases as $N_0 \propto R^{-2(\gamma + 2)/3}$ (to account for adiabatic losses) and the magnetic-field strength, $B$,
decreases as $B \propto R^{-1}$ (to account for energy conservation). In these expressions, energy gain by SSA is not taken into account. The maximum energy of the leptons, $E_{max}$, also decreases with increasing $R$, due to both adiabatic and radiative cooling.

In the launching (or concave) region, the jet diameter follows the relation $r \propto R^\beta$ (with $\beta$ positive and $\ll 1$), so it does not vary much with distance to the jet base. The bulk Lorentz factor, $\Gamma$, depends on distance as $\Gamma \propto R^\beta$ (i.e., the plasma is still being accelerated to be later injected into the conical jet region). A changing $\Gamma$ maps into a running Doppler shift and boost factor, which will steepen the observed spectrum. The particle density and the magnetic field change as $N_0 \propto R^{-\beta(\gamma + 2)}$ and $B \propto R^{-2\beta}$, respectively. The maximum cutoff energy in the lepton population, $E_{max}$, also decreases with increasing distance to the jet base.

\subsection{Implementation of the model}

Our simplified jet model makes use of the relationships described above, but with some simplifications. On the one hand, we assume that the cutoff energy, $E_{max}$, is always higher than the energy whose critical frequency corresponds to our highest observing frequency (see Appendix \ref{CoolApp} for a discussion on the implementation of lower high-energy cutoffs in the electron population.) 

On the other hand, we assume that the jet tube (where the plasma is confined) is very narrow, so that $r \ll R$ for all $R$. This way, the radiative-transfer equation can be solved easily in both, the conical and the concave jet region, as

\begin{equation}
I_\nu(R) = \delta(R)^3\frac{\epsilon_\nu(R)}{\kappa_\nu(R)}\left(1 - \exp{\left(-\frac{2\kappa_\nu(R)\,r}{\sin{\theta}} \right)}\right),
\label{InuEq}
\end{equation}

\noindent where $\delta$ is the Lorentz boost factor at a distance $R$ to the jet base, $\theta$ is the viewing angle of the jet (see Fig.\,\ref{SketchJet}); $\epsilon_\nu(R)$ is the synchrotron emissivity at frequency $\nu$ and distance $R$; and $\kappa_\nu(R)$ is the absorption coefficient. The opacity, $\tau_\nu$, is just $\kappa_\nu$ times the path length of the light-ray towards the observer, i.e., $2r/\sin{\theta}$). In this expression, we also assume that the opening angle of the tube, $\phi$, is much smaller than the viewing angle $\theta$, although this condition can be relaxed with no much change in the results \footnote{The main effect of a larger opening angle, $\phi$, as long as it does not approach $\theta/2$, is basically a change in the length of the light path within the jet, which can be rewritten as a change in the constant factor for $\tau_\nu$ in Eqs. \ref{EpKapCone} and \ref{EpKapLaunch}}. The quantities $\epsilon_\nu$ and $\kappa_\nu$ can be written in terms of $N_0$ and $B$ (e.g., \citealt{pac70}) in the form

\begin{equation}
\epsilon_\nu \propto N_0\,B^{(1+\gamma)/2}\,\left(\frac{\nu}{\delta}\right)^{(1-\gamma)/2}
\end{equation}

\noindent and

\begin{equation}
\kappa_\nu \propto N_0\,B^{(2+\gamma)/2}\,\left(\frac{\nu}{\delta}\right)^{-(4+\gamma)/2}.
\end{equation}

\noindent The two quantities, $B$ and $N_0$, are in turn power laws of $R$, so we can arrange the terms in all the power laws and write

\begin{equation}
\frac{\epsilon_\nu}{\kappa_\nu} \propto R^{1/2} \left(\frac{\nu}{\delta}\right)^{5/2}~~\textrm{and}~~\tau_\nu \propto R^{-(7\gamma+8)/6} \left(\frac{\nu}{\delta}\right)^{-(\gamma+4)/2},
\label{EpKapCone}
\end{equation}

\noindent for the conical region, and 

\begin{equation}
\frac{\epsilon_\nu}{\kappa_\nu} \propto R^\beta \left(\frac{\nu}{\delta}\right)^{5/2}~~\textrm{and}~~\tau_\nu \propto R^{-2\beta(\gamma+2)} \left(\frac{\nu}{\delta}\right)^{-(\gamma+4)/2},
\label{EpKapLaunch}
\end{equation}

\noindent for the concave region. Equation \ref{InuEq} can be solved, using Eqs. \ref{EpKapCone} or \ref{EpKapLaunch}, by imposing two boundary conditions to determine all the proportionality constants. The two conditions that we choose to solve Eq. \ref{InuEq} are, on the one hand, the value of $R$ for which we have a given opacity ($\tau=1$) at a given frequency (100\,GHz) and, on the other hand, the integrated flux density (i.e., the integral $F_\nu = \int{I_\nu\,d\Omega}$ over the jet) at a given frequency (100\,GHz). These quantities are given in table \ref{tab:parameters}. We notice that all the cosmological effects (e.g., redshift and time stretching) are taken into account and the absolute values of $B$ and $N_0$ are not needed in the model. 

The plasmon that originates the flare in our model is parametrized as an over-density in the electron population, of width $\Delta$ and contrast $K$. 
For a given location of the plasmon, $R'$, we make $N(R) \rightarrow K\,N(R)$ for $R \in [R'-\Delta/2,R'+\Delta/2]$. 

The magnetic field in the region of the plasmon is also scaled, to keep the ratio of particle energy density to magnetic-field energy density unchanged, with respect to that in the quiescent state. The magnetic field and the particle density are related as $B \propto N^\eta$. Hence, we scale $B$ with the factor $K^\eta$. The parameter $\eta$ takes the values $3/(2\gamma+4)$ (case of the conical region) and $2/(\gamma+2)$ (case of the concave region).

In our model, the plasmon travels through the jet at a speed determined by the Lorentz factor, $\Gamma$ (computed for each distance, $R$) by keeping $\Delta$ and $K$ constant. Effects of the finite light travel-time from the different $R$ to the observer are also taken into account.

\subsection{Known limitations of the model}

Besides the simplifications described in the previous section of this appendix, there are other limitations in the model that have to be noticed. In particular,

\begin{itemize}

\item We use an ad-hoc model for the plasmon and its evolution. The width and contrast factor, with respect to that of the quiescent plasma, is assumed to be constant. However, it may vary as the plasmon departs from the jet base. A more realistic model of plasmon will depend on the particular physical mechanisms related to its generation (e.g., a shock-shock interaction). Indeed, flaring activity can also be obtained by changing, not only $N_0$, but other parameters involved in the intensity of the jet emission (e.g., $\gamma$, $\delta$, $\Gamma_0$, etc.)

\item Evolving electron population due to radiative energy losses. Indeed, the mean life-time of the electrons could be short at the high critical frequencies of our observations. This, however, depends on several quantities, as the strength of the magnetic field, that depend, in turn, on the absolute diameter of the jet tube, which is undetermined in our model. 

\item The width of the plasmon shall be constrained by the cooling time of the electrons, which in turn depends on their energy.

\item A more accurate radiative transfer should take into account the diameter of the jet tube, the different values of $N_0$ and $B$ found during the path of the light ray, and the effects from the finite light-travel time through the width of the jet.

\item We assume a jet with smooth variation of magnetic field and particle density. We do not consider the presence of standing shocks close to the jet base.

\end{itemize}

\subsection{Effects of radiative and adiabatic cooling} \label{CoolApp}

\cite{mar80} models the effect of radiative and adiabatic cooling by using a maximum (i.e., cutoff) energy in the population of relativistic electrons. The cutoff energy decreases as a power law of the distance to the jet base, being the exponent dependent on several parameters related to adiabatic and synchrotron losses. In the scenario of a conical jet (i.e., peak intensity located far from the jet base, due to SSA effects) with a small viewing angle, an electron population with a high-energy cutoff is the only way to generate a steep spectrum similar to the one obtained in our observations. However, we notice that such an electron population would not be able to reproduce the changing flux ratios reported in this paper. 

The reason for this statement is subtle. The main contribution to the flux density at a given frequency comes from the region around the $\tau=1$ surface at that frequency (i.e., the VLBI core). Hence, a steep spectrum shall be obtained if, and only if, the cutoff energies corresponding to the higher critical frequencies are achieved in the jet region {\em behind} the core (i.e., at $R$ smaller than that one corresponding to the $\tau=1$ surface). Hence, a flare like that used in our model could never increase the flux density at the higher frequencies (hence changing the flux ratio between the lensed images), since the flux at these frequencies would always be similar to the source function (i.e., the emission at the optically-thick region, which is independent of $N_0$).

If we use instead an electron population with a smooth, although fast, energy decrease after a given critical energy, the problem discussed in the previous paragraph could, in principle, be solved. Let us use a population of electrons following the energy distribution (e.g., \citealt{pot12})

\begin{equation}
N(E) = N_0 E^{-\gamma}\,\exp{\left(-E/E_{max}\right)},
\label{newNE}
\end{equation}

\noindent where $E_{max}$ can be written as a power law of $R$. If we assume that an electron with energy $E$ radiates all its power at the critical frequency $\nu = C B\,E^2$ (where $C$ is a constant, e.g. \citealt{pac70}), it can be easily shown that the emission and absorption coefficients of the whole electron population are

\begin{equation}
\epsilon_\nu \propto N(E)\,\left.\left(B\,\frac{\nu}{\delta}\right)^{1/2}\right|_{E \rightarrow \left(\frac{\nu}{C\,\delta\,B}\right)^{1/2}}
\label{newEps}
\end{equation}

\noindent and

\begin{equation}
\kappa_\nu \propto -E^2\frac{d}{dE}\left(\frac{N(E)}{E^2}\right)\,\left.\left(B\,\frac{\nu}{\delta}\right)^{1/2}\right|_{E \rightarrow \left(\frac{\nu}{C\,\delta\,B}\right)^{1/2}}.
\label{newKap}
\end{equation}

We recall that $\delta$ is the Doppler boost factor. These new equations can be used to solve Eq. \ref{InuEq} using a generic electron population, $N(E)$. In our case, these equations reduce to

\begin{equation}
\epsilon_\nu \propto N_0 \left(\frac{\nu}{\delta}\right)^{(1-\gamma)/2}\exp{\left( \frac{-1}{E_{max}}\sqrt{\frac{\nu}{C\,\delta\,B}}\right)} B^{(1+\gamma)/2}
\label{newEps2}
\end{equation}

\noindent and

\begin{equation}
\kappa_\nu \propto N_0 \left(\frac{\nu}{\delta\,B}\right)^{-\gamma}\,\exp{\left( \frac{-1}{E_{max}}\sqrt{\frac{\nu}{C\,\delta\,B}}\right)}\,\left(\frac{(\delta\,B)^{1/2}(\gamma+2)}{\nu^{1/2}} + \frac{1}{E_{max}}\right).
\label{newKap2}
\end{equation}

We have applied Eqs. \ref{newEps2} and \ref{newKap2} to simulate a conical jet with radiative and/or adiabatic energy losses taken into account. In this case, $E_{max} \propto R^{-1}$. However, we have been unable to reproduce a spectrum as steep as that shown by the data, and in any case have we been able to reproduce the behaviour shown by the flux-density ratios as a function of frequency and time. Hence, we conclude that our data are incompatible with the scenario of a conical jet, as long as our model of the jet flare, used to explain the observed evolving flux-density ratios, holds. 

\end{document}